\documentclass[lengthcheck,unsortedaddress,superscriptaddress,showpacs,letterpaper,nofootinbib]{revtex4-1}

\usepackage{comment}
\usepackage{color}
\usepackage{graphicx}  
\usepackage{float}
\usepackage{amsmath}
\usepackage{subfig}
\usepackage{acronym}
\usepackage{appendix}
\usepackage[bookmarksnumbered, bookmarksopen, breaklinks, colorlinks]{hyperref}

\newcommand{\AEI}{Albert-Einstein-Institut, Max-Planck-Institut f\"ur
Gravitationsphysik, D-30167 Hannover, Germany}
\newcommand{\Leibniz}{Leibniz Universit\"at Hannover, D-30167 Hannover, Germany}
\newcommand{\PSU}{Institute for Gravitation and the Cosmos, Department of Physics, The Pennsylvania State University, University Park, PA 16802, USA}

\begin{document}

\title{Parameter space metric for 3.5 post-Newtonian gravitational-waves from compact binary inspirals}

\author{Drew~Keppel}  
\email{drew.keppel@ligo.org}
\affiliation{\AEI}
\affiliation{\Leibniz}

\author{Andrew~P.~Lundgren}  
\email{andrew.lundgren@ligo.org}
\affiliation{\AEI}
\affiliation{\Leibniz}
\affiliation{\PSU}

\author{Benjamin~J.~Owen}  
\email{ben.owen@ligo.org}
\affiliation{\PSU}

\author{Hanyuan Zhu}  
\email{hanyuan.zhu@ligo.org}
\affiliation{\PSU}

\begin{abstract}

We derive the metric on the parameter space of 3.5 post-Newtonian (3.5PN)
stationary phase compact binary inspiral waveforms for a single detector,
neglecting spin, eccentricity, and finite-body effects.  We demonstrate that
this leads to better template placement than the current practice of using the
2PN metric to place 3.5PN templates: The recovered event rate is improved by
about 10\% at a cost of nearly doubling the number of templates.  The
cross-correlations between mass parameters are also more accurate, which will
result in better coincidence tests.

\end{abstract}

\maketitle

\acrodef{BBH}{binary black hole}
\acrodef{BNS}{binary neutron star}
\acrodef{GW}{gravitational wave}
\acrodef{ISCO}{inner-most stable circular orbit}
\acrodef{NSBH}{neutron-star black-hole binary}
\acrodef{PN}{post-Newtonian}
\acrodef{PSD}{power spectral density}
\acrodef{SNR}{signal-to-noise ratio}
\acrodef{SPA}{stationary phase approximation}

\section{Introduction}

Many searches for \acp{GW} from coalescing compact binaries (neutron stars
and/or black holes) have been conducted in data from large-scale
interferometric detectors for more than a decade~\cite{Tagoshi:2000bz,
Takahashi:2004bj, CBC:S1, CBC:S2BNS, CBC:S2MACHO, CBC:S2BBH, CBC:LIGOTAMA,
CBC:S3S4, CBC:S3Spin, CBC:S51yr, CBC:S5:12_18, CBC:S5LV, CBCHM:S5, CBC:S6,
LIGO:CBC:S5:GRB, GRB070201, GRB051103, LIGO:S6:GRB}, and these searches will
remain key science goals in the era of advanced detectors and
beyond~\cite{Harry:2010zz}.  These searches achieve their sensitivity using
matched filtering, which relies on the availability of template waveforms that
well approximate the signals.  For example, the inspiral phase of the
coalescence, an adiabatic evolution of quasi-circular orbits which dominates
the waveform for most of the published searches, is well modeled by the \ac{PN}
approximation (see \cite{Blanchet:2006zz} for a review).  Here we focus on
these well-modeled waveforms and mostly neglect any discrepancies with the true
signals.

Even well-modeled waveforms have a parameter space to cover (for instance the
masses of the compact objects) as the \ac{SNR} degrades if the signal is
filtered with a template with the wrong parameters, even if the latter has the
correct functional form.  Therefore a key component of matched filtering
searches is the construction of template banks such that, no matter where in
the parameter space a signal appears, at least one template has parameters
close enough to achieve a predefined fraction of the optimal \ac{SNR} known as
the \textit{minimal match} (named in~\cite{Owen:1995tm}; a similar concept was
first mentioned in the \ac{GW} literature in~\cite{Sathyaprakash:1991mt}).  The
\textit{match} between two templates is derived from what is known in
statistics as the ambiguity function (e.g.~\cite{Helstrom}), maximizing the
latter over differences in \textit{extrinsic parameters} to account for
features of the detection algorithm such as maximization over a phase constant.
Thus the match depends only on \textit{intrinsic parameters} describing the
shapes of the waveforms, such as masses of the two components of a binary (the
distinction depends on the signal; see footnote~[15] of
Ref.~\cite{Owen:1995tm}).

Construction of a template bank with a given minimal match requires an
algorithm for placement of the templates in the parameter space.  Ideally, one
would cover the parameter space with the fewest templates given a fixed value
of minimal match so as to minimize the computational cost.  There are several
algorithms proposed or implemented~\cite{Owen:1995tm, Cokelaer2007, Prix2007,
Harry2009, Messenger2009, Manca2010}, involving either the construction of
regular lattices in the parameter space or stochastic point selection followed
by match-based selection criteria.  These algorithms have varying computational
costs and achieve varying approximations to the optimal (minimum) number of
templates for a fixed minimal match, generally trading off those features
versus robustness against parameter choices and varying dimensionality of
parameter space.

All template placement algorithms can benefit at some level from the meric on
the search parameter space.  This metric relates the loss of match between
adjacent templates to the proper distance between them in a simple and
computationally cheap way as long as the proper distance is not large.  For
template banks with large spacing (i.e.\ low minimal match) such as in
continuous-wave searches, the standard relation is less accurate but could be
made more accurate by using the Riemannian curvature derived from the metric.
The search metric has also been used in other aspects of data analysis. In
particular, this metric has been used to cluster single detector
triggers~\cite{Robinson2007} and to check whether triggers from different
detectors are coincident~\cite{Robinson:2008un}.  The search metric involves
only the intrinsic parameters, and is related to but distinct from the Fisher
information matrix, which has been used as a metric on the full parameter space
for decades outside the field of gravitational waves---see, e.g.,
Ref.~\cite{Helstrom}.

The two-dimensional metric for the inspiral portion of the signal associated
with \acp{GW} from \acp{BNS} was originally computed at 1.0 \ac{PN}
order~\cite{Owen:1995tm} and then extended to 2.0 \ac{PN}~\cite{Owen:1998dk}.
(The \ac{PN} order refers to the highest power beyond leading order of the
square of the orbital velocity kept in the waveform.) It was also noted that a
different choice of coordinates resulted in a slowly varying metric that, in
addition, had an analytic transformation to and from the mass coordinate
space~\cite{Tanaka:2000xy, Croce:2001wf, Babak2006}. Recent searches for
inspiral \ac{GW} signals from binaries with negligible spin have used the
latter form of this metric along with an optimal packing (hexagonal lattice)
template placement routine in two dimensions~\cite{Cokelaer2007}. This has been
case even when the waveforms themselves were changed to include the
3.5~\ac{PN}~\cite{CBC:S51yr} terms in the waveform phase.  To enhance the
sensitivity of a search at a fixed false alarm rate, it would be better to use
the metric computed to the same order as the waveforms.  The reason is that
there is a \textit{fitting factor} issue as well as a minimal match issue: The
former is a quantity similar to the latter, describing instead the loss of
signal-to-noise due to imperfect waveform models~\cite{Apostolatos:1995pj}.
Total loss of \ac{SNR} is bounded above by the sum of the losses; see e.g.\ the
discussion around Fig.~3 of Ref.~\cite{Lindblom:2008cm}.  Even at the same
\ac{PN} order there may be some remaining fitting factor issue if the templates
use a different type of approximant (for example, see~\cite{Buonanno:2009zt}
for how different approximants of 3.5 \ac{PN} order compare). We neglect this
and continue to focus on parameter-space discretization issues rather than
waveform modeling issues.

Hence in this work we extend the non-spinning inspiral metric of
\cite{Owen:1995tm} and \cite{Owen:1998dk} to 3.5~\ac{PN} order in phase.
Preliminaries of the computation can be found in Sec.~\ref{sec:prelim}. The
metric computation itself is found in Sec.~\ref{sec:metric} and validations of
it in Sec.~\ref{sec:comparison}. 

\section{Formalism}
\label{sec:prelim}

We use the \ac{SPA} inspiral waveforms, which are known to be accurate enough
for most purposes~\cite{Chassande1998, Droz:1999qx}.  The general \ac{SPA}
inspiral waveform can be written in the form~\cite{Sathyaprakash:1991mt}
\begin{align}\label{eq:waveform}
h(f) &= \frac{\mathsf{A}(t_f)}{\sqrt{\dot{F}(t_f)}} \mathrm{e}^{i \Psi(t_f)} \\
&= \mathcal{A}(f) \mathrm{e}^{i \Psi(f)},
\end{align}
where $\mathsf{A}(t)$ is the time-domain waveform amplitude, $\dot{F}(t)$ is
the instantaneous \ac{GW} frequency, and $t_f$ is defined to be the time where
$F(t_f) = f$ (the Fourier frequency).  The combined amplitude $\mathcal{A}(f)$
truncated at Newtonian order~\cite{Sathyaprakash:1991mt, CutlerFlanagan1994} is
given as
\begin{equation}
\mathcal{A}(f) = \sqrt{\frac{5}{24}} \frac{\mathcal{M}^{-5/3}}{\pi^{2/3} D}
f^{-7/6}.
\end{equation}
For convenience, we define the non-frequency dependent part of the amplitude to
be $A$, where
\begin{equation}\label{eq:amp}
A := \sqrt{\frac{5}{24}} \frac{\mathcal{M}^{-5/3}}{\pi^{2/3} D}.
\end{equation}
The phase up to 3.5~\ac{PN} order is given as~\cite{Arun:2004hn}
\begin{multline}\label{eq:phase}
\Psi(f) = 2 \pi t_c f - \phi_c - \pi/4 + \sum_{j=0}^{7} \psi_{\frac{j}{2}
\mathrm{PN}} f^{(-5+j)/3} \\
+ \sum_{j=5}^{6} \psi^l_{\frac{j}{2} \mathrm{PN}} \ln(f) f^{(-5+j)/3}
\end{multline}
where the \ac{PN} coefficients $\{\psi_{\frac{j}{2} \mathrm{PN}},
\psi^l_{\frac{j}{2} \mathrm{PN}}\}$ are given in Appendix~\ref{ap:PNcoeffs} in
terms of the symmetric mass ratio $\eta = m_1 m_2 / M^2$ and chirp mass
$\mathcal{M} = M \eta^{3/5}$, where $M = m_1 + m_2$ is the total mass of the
binary and $m_1$ and $m_2$ are the component masses.  These intrinsic
parameters affect the shape of the waveform and are fundamental to the search
problem, while the extrinsic parameters (here the coalescence time $t_c$ and
coalescence phase $\phi_c$) are easily searched over (see below).  The
combination of leading-order amplitude and higher-order phase is commonly
called the \textit{restricted PN} approximation~\cite{CutlerFlanagan1994}.

In searching for well modeled signals, such as the inspiral phase of signals
associated with compact binary coalescence in a \ac{GW} detector's data, the
optimal filter in the sense of maximizing \ac{SNR} is the matched filter.  To
compute the matched filter for a specific template waveform, we define the
inner-product between two time vectors $x$ and $y$ in the frequency domain
as~\cite{Finn:1992wt, CutlerFlanagan1994}
\begin{equation}\label{eq:innerproduct}
(x|y) := 4 \Re \int_0^\infty \frac{\tilde{x}(f) \tilde{y}^*(f)}{S_n(f)} df,
\end{equation}
where $S_n(f)$ is the single-sided \ac{PSD} associated with the noise of the
detector. With a time delay $t$ between the two vectors the convolution theorem
lets us write
\begin{equation}
(x|y)(t) := 4 \Re \int_0^\infty \frac{\tilde{x}(f) \tilde{y}^*(f)}{S_n(f)}
e^{-2 \pi i f t} df.
\end{equation}
In practice, the lower limit of this integral is chosen to be a point below
which a negligible amount of signal power is lost, mainly a function of a
detector's \ac{PSD}. The upper limit depends on the frequency band
for which the \ac{SPA} signal has support.  Ideally, matched filtering data $x$
with a single template waveform $h$ involves taking the inner product
\begin{equation}
\rho(t) = (x|h)(t).
\end{equation}
For detection of signals with unknown time of arrival this quantity is maximized
over $t$ ($t_c$ in the case of inspirals), which can be done quickly and simply
with a fast Fourier transform.  For sinusoidal signals there is generally an
unknown constant phase offset as well ($\phi_c$ in the case of inspirals),
which is dealt with computationally efficiently as follows: The square \ac{SNR}
associated with template waveform $h$ is the sum-of-squares combination of the
filter outputs of the 0 and $\pi/2$ phases of the waveform,
\begin{equation}
\rho^2(t) = \frac{(s|h_0)^2(t) + (s|h_{\pi/2})^2(t)}{(h_0|h_0)}.
\end{equation}
These parameters (and others which are quickly maximized over in searches for
other waveforms) are known as \textit{extrinsic parameters} as opposed to
\textit{intrinsic parameters} affecting the waveforms shape, and the
distinction affects the template placement problem as follows:

Consider the inner product between two waveforms
\begin{equation}
(h(\theta)|h(\theta+\Delta\theta)),
\end{equation}
where we use $h(\theta)$ to denote a waveform with parameter vector
$\theta^\alpha$.  This is known in the statistical literature as the
\textit{ambiguity function}~\cite{Helstrom}.  For small $\Delta\theta$ and
well-behaved parameters (except for the overall amplitude) the ambiguity
function falls off quadratically in $\Delta\theta$ from the maximum at
$\Delta\theta=0$,
\begin{equation} \label{eq:mismatch}
(h(\theta)|h(\theta+\Delta\theta)) \approx (h(\theta)|h(\theta)) -
\Gamma_{\alpha\beta} \Delta\theta^\alpha \Delta\theta^\beta,
\end{equation}
where the Fisher information matrix
\begin{equation}
\Gamma_{\alpha\beta} = (\partial_\alpha h|\partial_\beta h)
\end{equation}
acts as a metric on the entire parameter space.  (The partial derivatives are
with respect to the associated parameter, i.e., $\partial_\alpha :=
\partial/\partial \theta^\alpha$.)

What is needed for template banks is the metric on the intrinsic parameter
space.  This can be obtained from a modified ambiguity function called the
\textit{match} in the same way that the information matrix is obtained from the
ambiguity function.  Breaking the vector $\theta^\alpha$ into extrinsic
parameters $\mu^\alpha$ (such as the overall amplitude in general and $t_c$ and
$\phi_c$ for inspiral signals) and intrinsic parameters $\lambda^\alpha$ (such
as the two masses for simple inspiral signals), we have the match as
Eq.~\eqref{eq:innerproduct} normalized to unit maximum and maximized over
$\Delta\mu^\alpha$, and the intrinsic parameter
metric as the corresponding matrix of coefficients in the expansion for small
$\Delta\lambda$~\cite{Owen:1995tm}.

The intrinsic parameter metric can also be obtained by a series of projections
from the information matrix, of the form
\begin{equation}\label{eq:projection}
g_{\alpha \beta}' = g_{\alpha \beta} - g^{a b} g_{\alpha a} g_{b \beta},
\end{equation}
where indices $a$ and $b$ include the parameter projected out as well as the
other parameters in $\alpha$ and $\beta$. This is equivalent to the Schur
complement of the parameter subspace we are projecting out. This is the
approach we take below.

\section{Computing The Inspiral Metric}
\label{sec:metric}

We define the moments of the detector \ac{PSD} similarly to but not quite the
same as other papers~\cite{PoissonWill1995, Owen:1995tm, Owen:1998dk,
Babak2006}. We use
\begin{subequations}\label{eq:moments}\begin{align}
I(q,l) &:= \int_{f_L}^{f_U} \frac{f^{-q / 3} \ln^{l}(f)}{S_n(f)} df \\
J(q,l) &:= I(q,l) / I(7,0).
\end{align}\end{subequations}
The moment functional is
\begin{equation}\label{eq:normalizedmoments}
\mathcal{J}[a] := \frac{1}{I(7,0)} \int_{f_L}^{f_U} \frac{f^{-7/3}}{S_n(f)}
a(f) df
\end{equation}
which with our conventions yields
\begin{equation}\label{eq:normalizedmomentsexpanded}
\mathcal{J} \left[ \sum_q a_q f^{-q/3} \ln^l(f) \right] = \sum_q a_q J(q-7,l).
\end{equation}
The logarithms are necessary to deal with derivatives of the higher order
\ac{PN} waveforms.

Starting from the information matrix, we can first project out the amplitude
$A$ by normalizing the information matrix~\cite{PoissonWill1995}
\begin{equation}\label{eq:metricfisher}
g_{\mu \nu} := \frac{\Gamma_{\mu \nu}}{\rho^2}.
\end{equation}

Using the definition of the inner product~\eqref{eq:innerproduct}, the
decomposition of the waveform $h$~\eqref{eq:waveform}, and the definition of
the moment functional~\eqref{eq:normalizedmoments}, the normalized information
matrix takes the form
\begin{equation}\label{eq:fullmetric}
g_{\mu \nu} = \mathcal{J}\left[ \frac{\partial \ln \mathcal{A}}{\partial
\lambda^\mu} \frac{\partial \ln \mathcal{A}}{\partial \lambda^\nu} \right] +
\mathcal{J}\left[ \frac{\partial \Psi}{\partial \lambda^\mu} \frac{\partial
\Psi}{\partial \lambda^\nu} \right],
\end{equation}
which can be expanded using \eqref{eq:amp}, \eqref{eq:phase},
\eqref{eq:normalizedmomentsexpanded}, and \eqref{eq:moments} to
\begin{align}\label{eq:fullmetricexpanded}
g_{\mu \nu} &= \frac{\partial \ln A}{\partial \lambda^\mu}\frac{\partial \ln
A}{\partial \lambda^\nu}J(7,0) + \left( 4\pi^2 \right) \frac{\partial
t_c}{\partial \lambda^\mu} \frac{\partial t_c}{\partial \lambda^\nu} J(1, 0)
\nonumber \\
&+ \left( -2\pi \right) \left( \frac{\partial t_c}{\partial \lambda^\mu}
\frac{\partial \phi_c}{\partial \lambda^\nu} + \frac{\partial \phi_c}{\partial
\lambda^\mu} \frac{\partial t_c}{\partial \lambda^\nu}\right) J(4, 0) \nonumber
\\
&+ \left( 2\pi \right) \sum_{i} \left( \frac{\partial t_c}{\partial
\lambda^\mu} \frac{\partial \psi_{i}}{\partial \lambda^\nu} + \frac{\partial
\psi_{i}}{\partial \lambda^\mu} \frac{\partial t_c}{\partial \lambda^\nu}
\right) J(9 - i, 0) \nonumber \\
&+ \left( 2\pi \right) \sum_{i} \left( \frac{\partial t_c}{\partial
\lambda^\mu} \frac{\partial \psi^l_{i}}{\partial \lambda^\nu} + \frac{\partial
\psi^l_{i}}{\partial \lambda^\mu} \frac{\partial t_c}{\partial \lambda^\nu}
\right) J(9 - i, 1) \nonumber \\
&+ \frac{\partial \phi_c}{\partial \lambda^\mu} \frac{\partial \phi_c}{\partial
\lambda^\nu} J(7, 0) \nonumber \\
&+ \left( -1 \right) \sum_{i} \left( \frac{\partial \phi_c}{\partial
\lambda^\mu} \frac{\partial \psi_{i}}{\partial \lambda^\nu} + \frac{\partial
\psi_{i}}{\partial \lambda^\mu} \frac{\partial \phi_c}{\partial \lambda^\nu}
\right) J(12 - i, 0) \nonumber \\
&+ \left( -1 \right) \sum_{i} \left( \frac{\partial \phi_c}{\partial
\lambda^\mu} \frac{\partial \psi^l_{i}}{\partial \lambda^\nu} + \frac{\partial
\psi^l_{i}}{\partial \lambda^\mu} \frac{\partial \phi_c}{\partial \lambda^\nu}
\right) J(12 - i, 1) \nonumber \\
&+ \sum_{i,j} \frac{\partial \psi_{i}}{\partial \lambda^\mu} \frac{\partial
\psi_{j}}{\partial \lambda^\nu} J(17 - i - j, 0) \nonumber \\
&+ \sum_{i,j} \left( \frac{\partial \psi_{i}}{\partial \lambda^\mu}
\frac{\partial \psi^l_{j}}{\partial \lambda^\nu} + \frac{\partial
\psi^l_{i}}{\partial \lambda^\mu} \frac{\partial \psi_{j}}{\partial
\lambda^\nu}\right) J(17 - i - j, 1) \nonumber \\
&+ \sum_{i,j} \frac{\partial \psi^l_{i}}{\partial \lambda^\mu} \frac{\partial
\psi^l_{j}}{\partial \lambda^\nu} J(17 - i - j, 2),
\end{align}
where all sums run from $0$ to $7$ for 3.5PN waveforms.
This expansion in \eqref{eq:fullmetric} and \eqref{eq:fullmetricexpanded}
discards the information associated with boundary terms from the derivatives.
That is, derivatives of $f_L$ and $f_U$ are neglected.
For the signals we consider, these are generally negligible since in practice
$f_L$ is generally chosen to be a constant and $f_U$ only becomes low enough
to significantly affect the moment integrals for high masses, where the search
is carried out by different means.

The metric used to construct search template banks must project out the
extrinsic parameters $t_c$ and $\phi_c$.  This results in a two-dimensional
metric on the space of intrinsic parameter only, which we refer to as the
\emph{mass metric}.

In the notation of \cite{Babak2006}, the triple projected mass metric $g_{\mu
\nu}'''$ is given as
\begin{widetext} \begin{align} \label{eq:projmetric}
g_{\mu \nu}''' =&\; \sum_{i,j,k,l} \boldsymbol{\Psi}_{\mu i k}
\boldsymbol{\Psi}_{\nu j l} \left[ J(17-i-j,k+l) -
\frac{J(12-i,k)J(12-j,l)}{J(7,0)} \right. \nonumber\\
 &\left.\quad\quad- \frac{\left[J(7,0)J(9-i,k) -
J(4,0)J(12-i,k)\right]\left[J(7,0)J(9-j,l) -
J(4,0)J(12-j,l)\right]}{J(7,0)\left[J(1,0)J(7,0) - J(4,0)J(4,0)\right]}\right]
;
\end{align} \end{widetext}
here $\boldsymbol{\Psi}_{\mu i 0} := \partial \psi_i / \partial \lambda^\mu$
and $\boldsymbol{\Psi}_{\mu i 1} := \partial \psi^l_i / \partial \lambda^\mu$
are tensors of derivatives of \ac{PN} coefficients and again the sums run from
$0$ to $7$ for 3.5PN waveforms.

We compute the derivatives of the phase with respect to the mass parameters
$\lambda^\mu \in \{\mathcal{M}, \eta\}$.  The derivatives of the \ac{PN} coefficients with
respect to the mass parameters can be found in Appendix~\ref{ap:PNcoeffderivs}.

An alternative derivation of this metric can be obtained by using the
Tanaka-Tagoshi procedure~\cite{Tanaka:2000xy}. We first define a ``premetric",
\begin{equation}
\gamma_{\alpha \beta} = \mathcal{J} \left[ \frac{\partial \Psi(f)}{\partial
C^\alpha} \frac{\partial \Psi(f)}{\partial C^\beta} \right],
\end{equation}
where $\Psi(f)$ is defined in \eqref{eq:phase} and $C^\mu \in \left\{ \psi_{j},
\psi^l_{j}, t_c, \phi_c \right\}$. This premetric would be constant and
therefore flat in the case where $f_U$ is fixed. The metric of
\eqref{eq:projmetric} is then obtained through the use of a Jacobian that
transforms from the coordinates $C^\alpha$ to coordinates $\lambda^\mu$ and the
projection of the $t_c$ and $\phi_c$ dimensions. We have used this as a check
of our derivation.

It is in fact more desirable to plot results in terms of the so-called chirp
times~\cite{Sathyaprakash:1994nj} parameters, which are similar to $C^\mu$.
The coordinate transformation is accomplished through the use of the Jacobian,
\begin{equation} \label{eq:jacobian}
g_{\mu' \nu'} = J^{\mu}_{\nu'} g_{\mu \nu} J^{\mu}_{\nu'},
\end{equation}
where $J^{\mu}_{\mu'} := \partial \lambda^\mu / \partial \lambda^{\mu'}$.  The
chirp-times we are interested in are $\tau_0$ and $\tau_3$, defined as
\begin{subequations}\begin{align}
\tau_0 &:= \frac{5}{256 \left(\pi f_0\right)^{8/3} \mathcal{M}^{5/3}}, \\
\tau_3 &:= \frac{\pi}{8 \left(\pi f_0\right)^{5/3} \mathcal{M}^{2/3}
\eta^{3/5}},
\end{align}\end{subequations}
and are related to the phase parameters $\psi_0$ and $\psi_3$. Here $f_0$ is a
reference frequency that, if chosen to be the lower frequency cutoff due to the
noise curve, results in $\tau_0$ being approximately the length of the
Newtonian waveform and $\tau_3$ being approximately the amount the waveform is
shortened by the 1.5~\ac{PN} terms.  The relevant terms of the Jacobian to
transform from the mass space to the chirp-times space are
\begin{subequations}\begin{align}
\frac{\partial \mathcal{M}}{\partial \tau_0} &= \frac{-768}{25} \left(\pi f_0
\mathcal{M}\right)^{8/3}, \\
\frac{\partial \mathcal{M}}{\partial \tau_3} &= 0, \\
\frac{\partial \eta}{\partial \tau_0} &= \frac{512}{15} \left(\pi
f_0\right)^{8/3} \mathcal{M}^{5/3} \eta, \\
\frac{\partial \eta}{\partial \tau_3} &= \frac{-40}{3 \pi} \left(\pi
f_0\right)^{5/3} \mathcal{M}^{2/3} \eta^{8/5}.
\end{align}\end{subequations}

\section{Comparison to 2.0~\ac{PN} metric}
\label{sec:comparison}

We compare the 2.0~\ac{PN} and 3.5~\ac{PN} versions of the two quantities most
important to template placement, the square root of the determinant of the
metric and the eigenvectors of the metric.

\begin{figure}
\begin{center}
\includegraphics[]{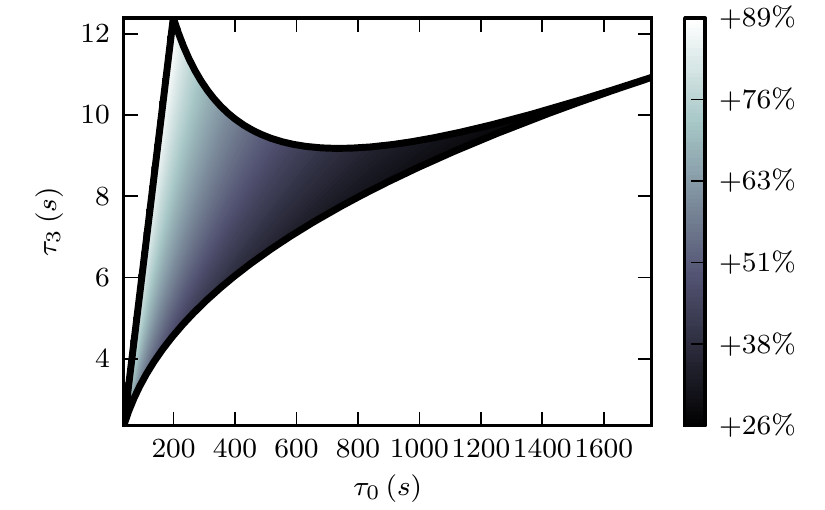}
\end{center}
\caption{Relative increase in template density $\sqrt{\left| g
\right|}$ when going from the 2.0~\ac{PN} mass metric to the 3.5~\ac{PN} mass
metric for the region of parameter space described in the text. The 3.5~\ac{PN}
mass metric density is between 25\% and 90\% larger, requiring that factor more
templates to cover the same (small) region of parameter space.}
\label{fig:density}
\end{figure}

Comparing the former quantity between the different \ac{PN} orders allows us to
see whether the density of template banks generated using the 2.0~\ac{PN}
metric is sufficient. The number of templates required is given
by~\cite{Owen:1995tm, Prix2007}
\begin{equation}\label{eq:numtemps}
N_{\rm templates} = \theta \left(m\right)^{-d/2} \int \sqrt{\left| g'''
\right|} d^d\lambda
\end{equation}
where $\theta$ is a geometrical quantity associated with how the template bank
tiles the parameter space, $m$ is the maximum mismatch allowed in the template
bank, $d$ is the dimensionality of the parameter space being tiled (i.e., two
for the present case as the templates are laid out in the non-spinning two
dimensional chirp-time space), and $\sqrt{\left| g''' \right|}$ is the square
root of the absolute value of the determinant of the metric on the intrinsic
parameter space.

Each template ``covers'' an elliptical region in intrinsic parameter space in
the sense that any other waveform within that ellipse will have a mismatch with
that template of no more than the (small) maximum mismatch $m$.  (Note that we
have switched from the previous notation of $m$ for mismatch between arbitrary
templates.) The elliptical shape is due to the quadratic approximation to the
mismatch in \eqref{eq:mismatch}, i.e.\ the approximation that all the
information on the mismatch is contained in the metric rather than higher
derivatives.  When this approximation holds, the principal axes of the
elliptical contours of constant mismatch are determined by the eigenvectors of
the metric~\cite{Owen:1995tm}.  Depending on the waveform family, for
mismatches of more than a few percent the quadratic approximation degrades and
the constant-mismatch contours acquire more complicated shapes than the metric
ellipses.

We compare the ratio of the metric densities between the 2.0~\ac{PN} metric and
the 3.5~\ac{PN} metric in Fig.~\ref{fig:density} for inspiral signals
associated with component masses between $1 M_{\odot}$ and $10 M_{\odot}$,
assuming an Advanced LIGO \ac{PSD} with the zero-detuning, high power
configuration~\cite{advLIGO}, a lower frequency cutoff of 10Hz, and an upper
frequency cutoff as the frequency associated with the \ac{ISCO} of the
Schwarzschild spacetime. (The latter is the cutoff commonly used in the
literature to approximate the division between the post-Newtonian inspiral and
fully relativistic merger phases of coalescence.) We see that the 3.5~\ac{PN}
metric has a density between 25\% and 90\% larger than the 2.0~\ac{PN} metric,
with the largest effect occurring along the left-most edge of the parameter
space, which corresponds to binaries with the larger mass being equal to $10
M_{\odot}$. This effect alone implies that template banks of 3.5~\ac{PN}
waveforms with a minimal-match of 97\% laid out using the 2.0~\ac{PN} metric
would actually only achieve a minimal-match of between $\sim$96\% and
$\sim$94\%. In other words, use of the 2.0~\ac{PN} template bank costs up to
$0.97^3 - 0.94^3 \sim$10\% of the ideal detection rate.

\begin{figure}
\begin{center}
\includegraphics[]{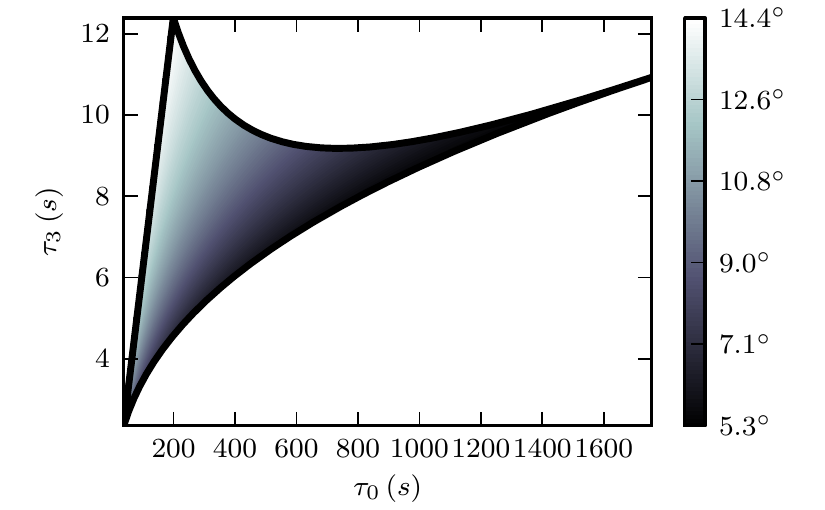}
\end{center}
\caption[Orientation Comparison]{Relative rotation of the
eigenvectors when going from the 2.0~\ac{PN} mass metric to the 3.5~\ac{PN}
mass metric for the region of parameter space described in the text. The
largest rotations are associated with binaries with the most unequal component
masses.}
\label{fig:orientation}
\end{figure}

In addition to the metric density, the eigenvectors of the metric (principal
axes of the metric ellipses) also change when going from the 2.0~\ac{PN} order
to 3.5~\ac{PN} order.  For the same parameter space as above, this can be seen
in Fig.~\ref{fig:orientation}. The largest rotation of the eigenvectors occurs
at the upper-left-most corner of the parameter space, which corresponds to the
binaries with the most asymmetric masses (i.e., $m_1 = 3 M_{\odot}$, $m_2 = 1
M_{\odot}$).

\begin{figure}
\centering
\subfloat[]{\includegraphics[]{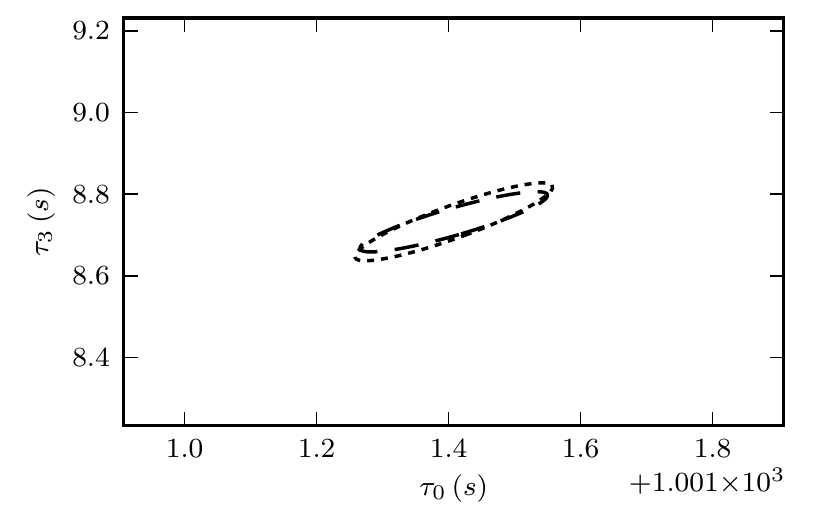}\label{fig:BNSmetrics}}
\qquad
\subfloat[]{\includegraphics[]{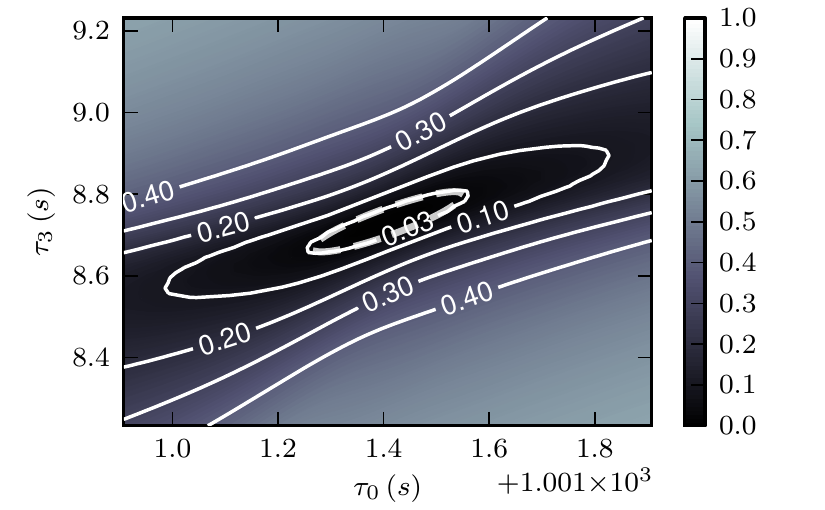}\label{fig:BNSmetriccheck}}
\caption[BNS Metric Comparison]{\subref{fig:BNSmetrics} The 2.0~\ac{PN}
(dotted) and 3.5~\ac{PN} (dashed) metric ellipses are compared for a \ac{BNS}
signal with $m_1=m_2=1.4 M_{\odot}$. \subref{fig:BNSmetriccheck} The
3.5~\ac{PN} metric ellipse (dashed) is plotted with a size corresponding to 3\%
loss of \ac{SNR}. The numerically computed full mismatch is shown in grey-scale
with certain contours denoted by solid lines. The metric ellipse matches the
3\% loss of \ac{SNR} contour; for higher mismatches the contours are not
elliptical.}
\label{fig:BNSmetric}
\end{figure}

\begin{figure}
\centering
\subfloat[]{\includegraphics[]{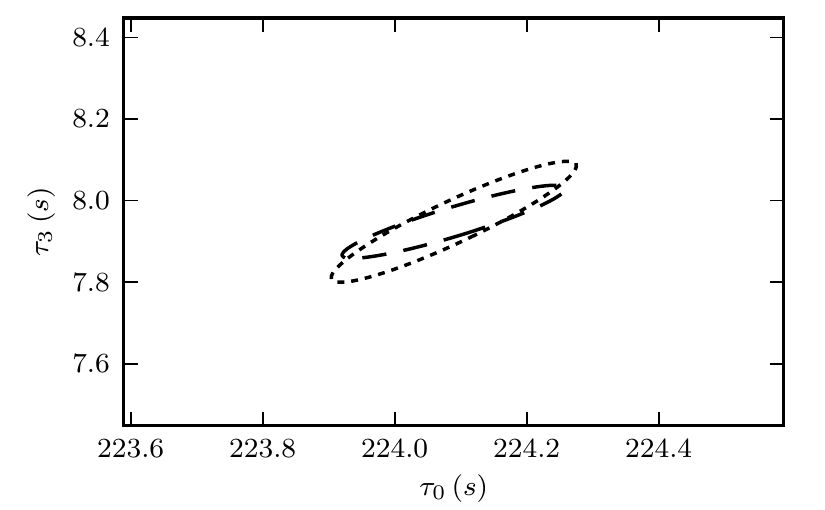}\label{fig:NSBHmetrics}}
\qquad
\subfloat[]{\includegraphics[]{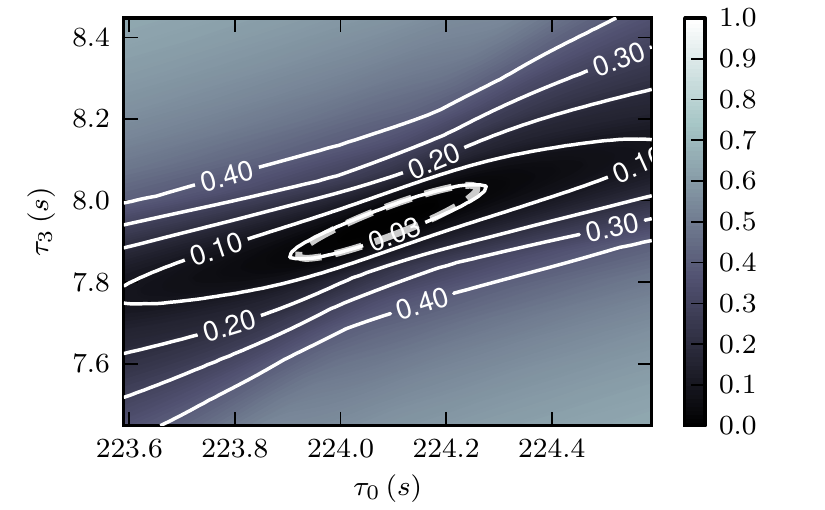}\label{fig:NSBHmetriccheck}}
\caption[NSBH Metric Comparison]{\subref{fig:NSBHmetrics} The 2.0~\ac{PN}
(dotted) and 3.5~\ac{PN} (dashed) metric ellipses are compared for a \ac{NSBH}
signal with $m_1= 10 M_{\odot}$ and $m_2=1.4 M_{\odot}$.
\subref{fig:NSBHmetriccheck} The 3.5~\ac{PN} metric ellipse (dashed) is plotted
with a size corresponding to 3\% loss of \ac{SNR}. The full numerical mismatch
is shown in grey-scale with certain contours denoted by solid lines. The metric
ellipse and the 3\% loss of \ac{SNR} fitting factor contour match less well
than in Fig.~\ref{fig:BNSmetriccheck}.}
\label{fig:NSBHmetric}
\end{figure}

\begin{figure}
\centering
\subfloat[]{\includegraphics[]{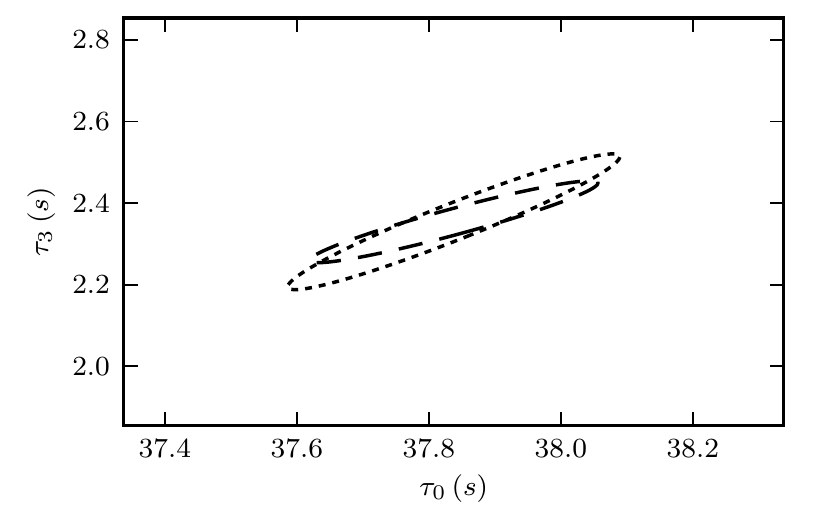}\label{fig:BBHmetrics}}
\qquad
\subfloat[]{\includegraphics[]{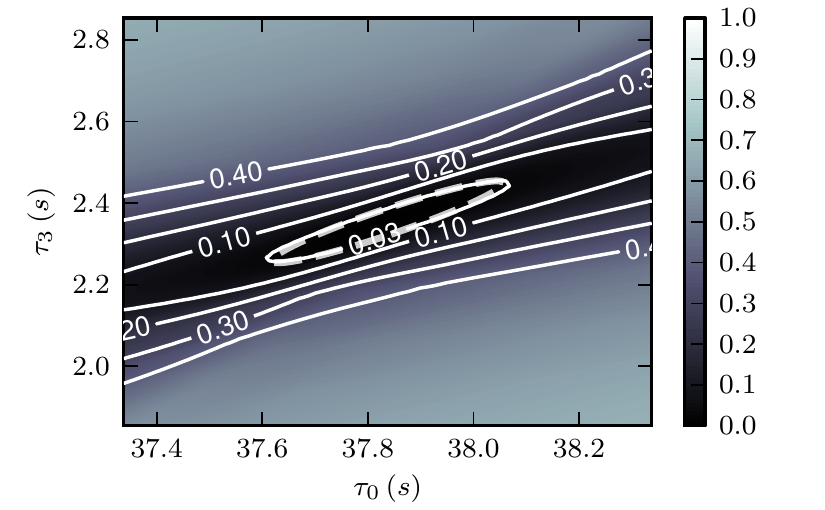}\label{fig:BBHmetriccheck}}
\caption[BBH Metric Comparison]{\subref{fig:BBHmetrics} The 2.0~\ac{PN}
(dotted) and 3.5~\ac{PN} (dashed) metric ellipses are compared for a \ac{BBH}
signal with $m_1=m_2=10 M_{\odot}$. \subref{fig:BBHmetriccheck} The 3.5~\ac{PN}
metric ellipse (dashed) is plotted with a size corresponding to 3\% loss of
\ac{SNR}. The full numerical mismatch is shown in grey-scale with certain
contours denoted by solid lines. The metric ellipse and the 3\% loss of
\ac{SNR} fitting factor contour match less well than in either
Fig.~\ref{fig:BNSmetriccheck} or \ref{fig:NSBHmetriccheck}.}
\label{fig:BBHmetric}
\end{figure}

These two effects are further visualized in Figs.~\ref{fig:BNSmetrics},
\ref{fig:NSBHmetrics}, and \ref{fig:BBHmetrics} for three points of interest in
the parameter space, the points associated with \ac{BNS}, \ac{NSBH} and
\ac{BBH}, respectively, where the mass of the neutron star is a canonical 1.4
$M_{\odot}$ and the mass of the stellar mass black hole is a canonical 10
$M_{\odot}$.  There we see metric ellipses associated with both the 2.0~\ac{PN}
and 3.5~\ac{PN} mass metric. The 2.0~\ac{PN} and 3.5~\ac{PN} metric ellipses
associated with \ac{NSBH} and \ac{BBH} signals disagree more than metric
ellipses associated with \ac{BNS} signals.  This is to be expected since, for a
given noise curve or frequency, waveforms of higher mass systems are more
influenced by higher-order \ac{PN} effects.

We also compare the 3.5~\ac{PN} metric ellipses with full mismatch contours in
Figs.~\ref{fig:BNSmetriccheck}, \ref{fig:NSBHmetriccheck}, and
\ref{fig:BBHmetriccheck}, again for points associated with \ac{BNS}, \ac{NSBH}
and \ac{BBH} systems, respectively.  Here ``full'' means not using the
quadratic approximation but rather numerically computing the mismatch using
\texttt{pylal}~\cite{lalsuite}. The \ac{BNS} metric ellipse associated with 3\%
loss of \ac{SNR} agrees well with the 3\% full mismatch contour. The same is
not true for the \ac{NSBH} and \ac{BBH} regions of parameter space. This could
be because the upper frequency cutoff for signals changes as a function of the
total mass of the system, which we neglect in our derivatives of the signal in
\eqref{eq:metricfisher}.

\begin{figure}
\centering
\subfloat[]{\includegraphics[]{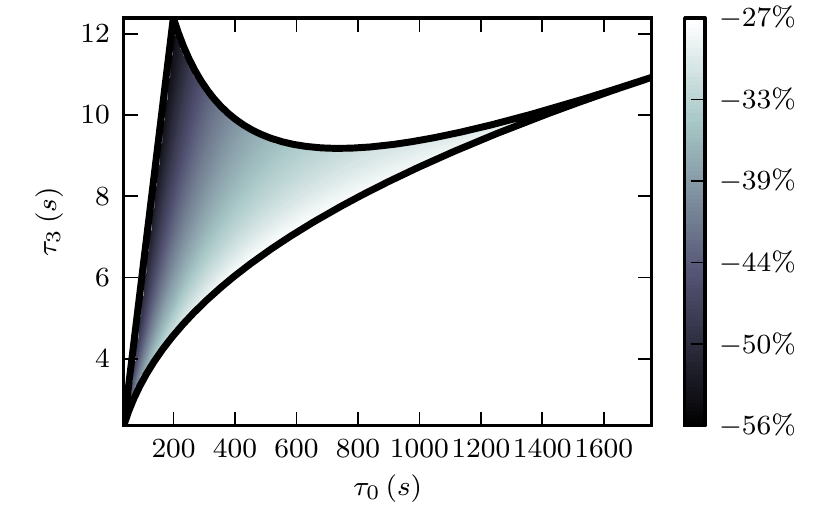}\label{fig:hexden}}
\qquad
\subfloat[]{\includegraphics[]{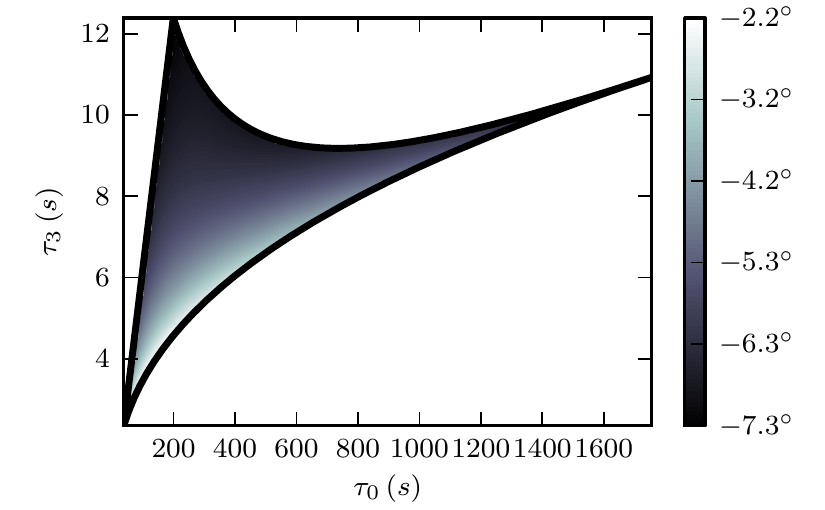}\label{fig:hexor}}
\caption[]{\subref{fig:hexden} The fractional difference between the
3.5~\ac{PN} mass metric density and the density of a 3.5~\ac{PN} ``fixed-mass''
metric, i.e.\ one computed at the most mass-asymmetric point with the upper
frequency cutoff set to be the Nyquist frequency, as implemented in the
standard template placement algorithm~\cite{Cokelaer2007}.  The fixed-mass
metric overcovers the entire parameter space.  \subref{fig:hexor} The
difference between the metric eigenvectors and those of the fixed mass metric.
The eigenvectors of the mass metric at 3.5~\ac{PN} are more aligned with those
of the 3.5~\ac{PN} fixed mass metric than with those of the 2.0~\ac{PN} metric
that varies as a function of parameter space.}
\label{fig:hex3p5}
\end{figure}

\begin{figure}
\centering
\subfloat[]{\includegraphics[]{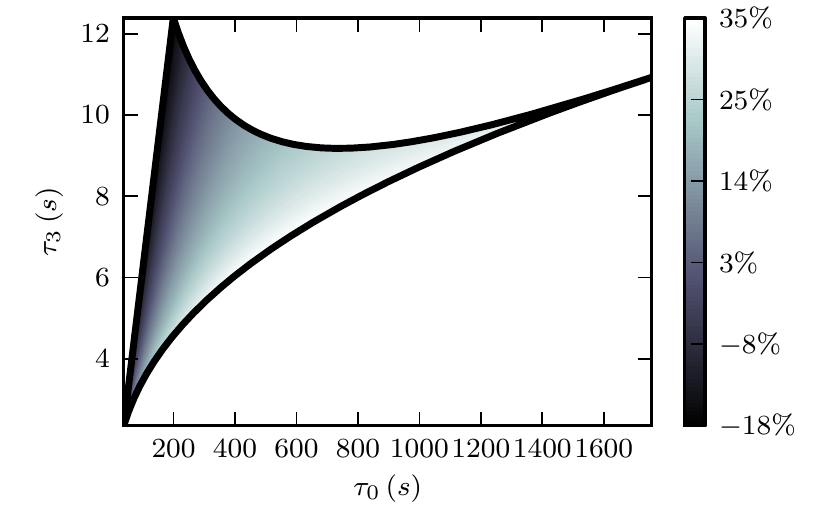}\label{fig:hexden2}}
\qquad
\subfloat[]{\includegraphics[]{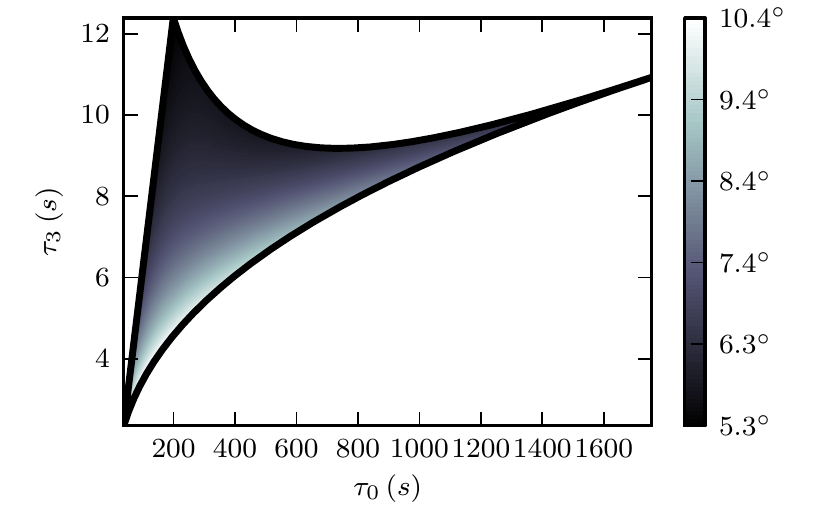}\label{fig:hexor2}}
\caption[]{\subref{fig:hexden2} The fractional difference between the
3.5~\ac{PN} mass metric density and the density of a 2.0~\ac{PN} ``fixed-mass''
metric, i.e.\ one computed at the most mass-asymmetric point with the upper
frequency cutoff set to be the Nyquist frequency, as implemented in the
standard template placement algorithm~\cite{Cokelaer2007}.  The fixed-mass
metric over and under covers portions of the parameter space.
\subref{fig:hexor2} The difference between the 3.5~\ac{PN} metric eigenvectors
and those of the 2.0~\ac{PN} fixed mass metric.  The eigenvectors of the mass
metric at 3.5~\ac{PN} are even less aligned with those of the 2.0~\ac{PN} fixed
metric than with those of the 3.5~\ac{PN} fixed metric.}
\label{fig:hex2p0}
\end{figure}

Now let us look at the total number of templates.  Let us use
\eqref{eq:numtemps} with $\theta = 2/3^{3/2}$, which is exactly true for the
optimal hexagonal lattice tiling of a flat two dimensional parameter space.
Performing the integral in \eqref{eq:numtemps} numerically for the region of
parameter space described above, we find that the 2.0~\ac{PN} mass metric
predicts that we will need $1.3 \times 10^5$ templates to cover the parameter
space, whereas the 3.5~\ac{PN} mass metric predicts $2.0 \times 10^5$
templates. However, the hexagonal template placement algorithm utilized for
previous searches \cite{Cokelaer2007} does not integrate the changing density.
Rather it measures the metric at the most asymmetric corner of the parameter
space, assumes the mass metric is invariant across parameter space, and uses
the Nyquist frequency as the high frequency cutoff for the noise moments. The
3.5~\ac{PN} metric density at this point is 86\% larger than that of the
2.0~\ac{PN} metric. If we use these point estimates in parameter space
integral, this results in $1.7 \times 10^5$ and $3.2 \times 10^5$ templates for
the 2.0~\ac{PN} and 3.5~\ac{PN} metrics, respectively.  As the metric actually
does vary throughout the parameter space, and as waveforms are only filtered up
to the frequency associated with the \ac{ISCO}, this tiling will use too many
templates. This can be seen in Fig.~\ref{fig:hex3p5} where \subref{fig:hexden}
shows the ratio of the 3.5~\ac{PN} mass metric density, which varies as a
function of parameter space, to the density of the ``fixed-mass'' metric
computed at the most asymmetric point as in~\cite{Cokelaer2007}.
Figure~\ref{fig:hexor} shows that the orientations of the eigenvectors of the
3.5~\ac{PN} metric that varies in parameter space are better aligned with those
of the 3.5~\ac{PN} fixed mass metric than with those of the 2.0~\ac{PN} metric
that varies in parameter space. Figure~\ref{fig:hex2p0} shows the same, expect
replacing the 3.5~\ac{PN} fixed metric with the 2.0~\ac{PN} fixed metric.
Although the 3.5~\ac{PN} fixed metric over covers the parameter space, for
constructing template banks, this is the more desirable side to err on. The
orientations of the eigenvectors of the 3.5~\ac{PN} fixed metric are also more
aligned with the correct 3.5~\ac{PN} metric than the orientations of the
eigenvectors of the 2.0~\ac{PN} fixed metric.

\section{Conclusion}

We have computed the inspiral metric associated with stationary phase waveforms
to 3.5~\ac{PN} order, extending the previously available metric at 2.0~\ac{PN}
order.  We have shown that the metric approximates the match well for
3.5~\ac{PN} waveforms for mismatch values of 3\% typically used in searches. We
have also characterized the error that is introduced by using a fixed metric
for the entire parameter space, as in~\cite{Cokelaer2007}. When a 2.0~\ac{PN}
fixed metric is used, some portions of the parameter space are under covered
and some over covered compared to the target minimal match. Using a 3.5~\ac{PN}
fixed metric causes the entire parameter space to be overcovered, however more
uniformly, a desirable feature for the template bank.  Using either a fixed
metric or a variable metric, the 3.5~\ac{PN} template bank requires nearly
twice as many templates as the equivalent 2~\ac{PN} bank.  Using only the
change in the density of the metric, we estimate that the 3.5~\ac{PN} template
bank reduces the loss of ideal event rate for a typical 3\%-mismatch template
bank to 10\% from about 20\% characterizing a 2~\ac{PN} template bank with
3.5~\ac{PN} waveforms.  The change in the parameter space metric, although it
changes the number of templates, does not change the false alarm rate for a
fixed \ac{SNR} threshold.  This is because it is the effective
\textit{independent} number of templates (the waveforms that are actually used
for the search) that affects the FAR, and it does so very weakly at realistic
thresholds.

In addition, the 3.5~\ac{PN} fixed metric more accurately reflects the
orientation of the eigenvectors of the 3.5~\ac{PN} metric. These results will
improve coincidence test and error estimates of inspiral \ac{GW} searches once
the advanced detectors become operational as they reflect a more accurate
understanding of how the parameters are correlated.

\acknowledgments

This work was supported by the Max Planck Gesellschaft and by National Science
Foundation grants PHY-0855589 and PHY-1206027. Numerical fitting factor
calculations in this work were accelerated using pycuda~\cite{pycudapyopencl}.
This document has LIGO document number LIGO-P1200165.

\appendix

\section{\ac{PN} Coefficients}
\label{ap:PNcoeffs}

The \ac{PN} coefficients $\psi_i$ and $\psi^l_i$ associated with the phase for
different \ac{PN} orders, where $i$ is twice the \ac{PN} order, are defined in
terms of the symmetric mass ratio $\eta$ and the chirp mass
$\mathcal{M}$,
\begin{equation}
\psi_{0} := \frac{3}{128 \pi^{5/3} \mathcal{M}^{5/3}},
\end{equation}
\begin{equation}
\psi_{2} := \frac{5}{384 \pi \mathcal{M} \eta^{2/5}} \left(\frac{743}{84} + 11
\eta\right),
\end{equation}
\begin{equation}
\psi_{3} := \frac{-3 \pi^{1/3}}{8 \mathcal{M}^{2/3} \eta^{3/5}},
\end{equation}
\begin{multline}
\psi_{4} := \frac{5}{3072 \pi^{1/3} \mathcal{M}^{1/3} \eta^{4/5}} \\ \times
\left(\frac{3058673}{7056} + \frac{5429}{7} \eta + 617 \eta^2 \right),
\end{multline}

\begin{multline}
\psi_{5} := \frac{5 \pi}{384 \eta} \left(\frac{7729}{84} - 13 \eta\right) \\
\times \left[1 + \ln\left(6^{3/2} \pi \mathcal{M} \eta^{-3/5}\right)\right],
\end{multline}
\begin{equation}
\psi^l_{5} := \frac{5 \pi}{384 \eta} \left(\frac{7729}{84} - 13 \eta\right),
\end{equation}
\begin{multline}
\psi_{6} := \frac{\pi^{1/3} \mathcal{M}^{1/3}}{128 \eta^{6/5}}
\left(\frac{11583231236531}{1564738560} - 640 \pi^2 \right. \\
\left.- \frac{6848}{7}\left[\gamma + \ln\left(4 \pi^{1/3} \mathcal{M}^{1/3}
\eta^{-1/5}\right)\right] \right. \\
\left.+ \frac{5}{4}\left[\frac{-3147553127}{254016} + 451 \pi^2\right] \eta
\right. \\
\left.+ \frac{76055}{576} \eta^2 - \frac{127825}{432} \eta^3\right),
\end{multline}
\begin{equation}
\psi^l_{6} := \frac{-107 \pi^{1/3} \mathcal{M}^{1/3}}{42 \eta^{6/5}},
\end{equation}
\begin{multline}
\psi_{7} := \frac{5 \pi^{5/3} \mathcal{M}^{2/3}}{32256 \eta^{7/5}} \\
\times \left(\frac{15419335}{336} + \frac{75703}{2} \eta - 14809 \eta^2\right).
\end{multline}
Any \ac{PN} coefficients of 3.5~\ac{PN} or lower not defined above are
identically zero. Compare Ref.~\cite{Arun:2004hn} where these were first
derived.

\section{Derivatives of \ac{PN} Coefficients}
\label{ap:PNcoeffderivs}

Here we give explicit expressions for the derivatives of the \ac{PN}
coefficients associated with the phase for different \ac{PN} orders in terms of
the symmetric mass ratio $\eta$ and the chirp mass $\mathcal{M}$. First the
derivatives with respect to $\mathcal{M}$,
\begin{equation}
\partial_{\mathcal{M}} \psi_{0} = \frac{-5}{128 \pi^{5/3} \mathcal{M}^{8/3}},
\end{equation}
\begin{equation}
\partial_{\mathcal{M}} \psi_{2} = \frac{-5}{384 \pi \mathcal{M}^2 \eta^{2/5}}
\left(\frac{743}{84} + 11 \eta\right),
\end{equation}
\begin{equation}
\partial_{\mathcal{M}} \psi_{3} = \frac{\pi^{1/3}}{4 \mathcal{M}^{5/3}
\eta^{3/5}},
\end{equation}
\begin{multline}
\partial_{\mathcal{M}} \psi_{4} = \frac{-5}{9216 \pi^{1/3} \mathcal{M}^{4/3}
\eta^{4/5}} \\ \times \left(\frac{3058673}{7056} + \frac{5429}{7} \eta + 617
\eta^2 \right),
\end{multline}
\begin{equation}
\partial_{\mathcal{M}} \psi_{5} = \frac{5 \pi}{384 \mathcal{M} \eta}
\left(\frac{7729}{84} - 13 \eta\right),
\end{equation}
\begin{multline}
\partial_{\mathcal{M}} \psi_{6} = \frac{\pi^{1/3}}{384 \mathcal{M}^{2/3}
\eta^{6/5}} \left(\frac{10052469856691}{1564738560} \right.  \\
\left.- 640 \pi^2 - \frac{6848}{7}\left[\gamma + \ln\left(4 \pi^{1/3}
\mathcal{M}^{1/3} \eta^{-1/5}\right)\right] \right. \\
\left.+ \frac{5}{4}\left[\frac{-3147553127}{254016} + 451 \pi^2\right] \eta
\right. \\
\left.+ \frac{76055}{576} \eta^2 - \frac{127825}{432} \eta^3\right),
\end{multline}
\begin{equation}
\partial_{\mathcal{M}} \psi^l_{6} = \frac{-107 \pi^{1/3}}{126 \mathcal{M}^{2/3}
\eta^{6/5}},
\end{equation}
\begin{multline}
\partial_{\mathcal{M}} \psi_{7} = \frac{5 \pi^{5/3}}{48384 \mathcal{M}^{1/3}
\eta^{7/5}} \\
\times \left(\frac{15419335}{336} + \frac{75703}{2} \eta - 14809 \eta^2\right).
\end{multline}
Now the derivatives with respect to $\eta$,
\begin{equation}
\partial_{\eta} \psi_{2} = \frac{-1}{384 \pi \mathcal{M} \eta^{7/5}}
\left(\frac{743}{42} - 33 \eta\right),
\end{equation}
\begin{equation}
\partial_{\eta} \psi_{3} = \frac{9 \pi^{1/3}}{40 \mathcal{M}^{2/3} \eta^{8/5}},
\end{equation}
\begin{multline}
\partial_{\eta} \psi_{4} = \frac{-3}{3072 \pi^{1/3} \mathcal{M}^{1/3}
\eta^{9/5}} \\ \times \left(\frac{3058673}{5292} - \frac{5429}{21} \eta + 1234
\eta^2 \right),
\end{multline}

\begin{multline}
\partial_{\eta} \psi_{5} = \frac{-\pi}{384 \eta^2} \\ \times
\left(\frac{7729}{84} \left[8 + 5 \ln\left(6^{3/2} \pi \mathcal{M}
\eta^{-3/5}\right)\right] - 39 \eta\right),
\end{multline}
\begin{equation}
\partial_{\eta} \psi^l_{5} = \frac{-38645 \pi}{32256 \eta^2},
\end{equation}
\begin{multline}
\partial_{\eta} \psi_{6} = \frac{-\pi^{1/3} \mathcal{M}^{1/3}}{640 \eta^{11/5}}
\left(\frac{11328104339891}{260789760} - 3840 \pi^2 \right. \\
\left.- \frac{41088}{7}\left[\gamma + \ln\left(4 \pi^{1/3} \mathcal{M}^{1/3}
\eta^{-1/5}\right)\right] \right. \\
\left.+ \frac{5}{4}\left[\frac{-3147553127}{254016} + 451 \pi^2\right] \eta
\right. \\
\left.- \frac{76055}{144} \eta^2 + \frac{127825}{48} \eta^3\right),
\end{multline}
\begin{equation}
\partial_{\eta} \psi^l_{6} = \frac{107 \pi^{1/3} \mathcal{M}^{1/3}}{35
\eta^{11/5}},
\end{equation}
\begin{multline}
\partial_{\eta} \psi_{7} = \frac{-\pi^{5/3} \mathcal{M}^{2/3}}{32256
\eta^{12/5}} \\
\times \left(\frac{15419335}{48} + 75703\eta + 44427 \eta^2\right).
\end{multline}
Any derivatives of \ac{PN} coefficients of 3.5PN or lower not defined above are
identically zero.

\bibliography{references}
\end{document}